\documentclass[12pt,a4paper]{article}
\pdfoutput=1
\usepackage[utf8]{inputenc}
\usepackage[affil -it]{authblk}

\usepackage{graphicx}
\usepackage{verbatim}
\usepackage{amsmath,amsfonts,amssymb,textcomp}
\usepackage{color}
\usepackage{authblk}

\begin{document}

\title{The current status of "Troitsk nu-mass" experiment in search for sterile neutrino}
\author[1]{D.\,N.\,Abdurashitov}
\author[1]{A.\,I.\,Belesev}
\author[1]{A.\,I.~Berlev}
\author[1]{ V.\,G.~Chernov}
\author[1]{ E.\,V.\,Geraskin}
\author[1]{A.\,A.\,Golubev}
\author[1,2]{G.\,A.~Koroteev}
\author[1]{N.\,A.~Likhovid}
\author[1]{A.\,A.~Lokhov}
\author[1]{A.\,I.~Markin}
\author[1,2]{A.\,A.~Nozik}
\author[1]{V\,.I.~Parfenov}
\author[1]{A.\,K.~Skasyrskaya}
\author[1]{V.\,S.\,Pantuev}
\author[1]{N.\,A.~Titov}
\author[1]{I.\,I.~Tkachev}
\author[1]{ F.\,V.~Tkachov}
\author[1]{S.\,V.~Zadorozhny}
\affil[1]{Institute for Nuclear Research of Russian Academy of Sciences, Moscow, Russia}
\affil[2]{Moscow Institute of Physics and Technology, Dolgoprudny, Russia}

\maketitle

\begin{abstract}
We propose a new experiment to search for a sterile neutrino in a few keV mass range at the "Troitsk nu-mass" facility. 
The expected signature corresponds to a kink in the electron energy spectrum in tritium beta-decay. The new goal compared to our previous experiment will be precision spectrum measurements well below end point. The experimental installation consists of a windowless gaseous tritium source and a high resolution electromagnetic spectrometer. We estimate that the current bounds on the sterile neutrino mixing parameter can be improved by an order of magnitude in the mass range under 5 keV without major upgrade of the existing equipment. 
Upgrades of calibration, data acquisition and high voltage systems will allow to improve the bounds by another order of magnitude. 

\end{abstract}


\section{Introduction}

The determination of the absolute mass scale and the number of neutrino mass eigenstates is a fundamental task both for particle physics, cosmology and astrophysics.
A non-zero mass for left-handed active neutrinos, indirectly observed in the neutrino oscillation experiments, suggests existence of right-handed sterile neutrinos.
The range of possible neutrino mass values for right-handed neutrinos is unrestricted at present.  
The results of short base-line oscillation experiments such as LSND~\cite{Aguilar:2001ty} or MiniBooNE~\cite{Aguilar-Arevalo:2013pmq}, the reactor anomaly~\cite{Mention:2011rk} and the calibration analysis of solar neutrino radiochemical experiments~\cite{Giunti:2010zu} can be explained with an assumption of one or two light sterile neutrinos in the mass range of few eV. 
The assumption of one very light neutrino state in addition to the three active states seems controversial within the standard cosmology, but may be compatible with the modern cosmological data. 
On the other hand a sterile neutrino in the keV mass range is a natural candidate for the dark matter. 
Similarly to the three generations of fermions in the Standard Model, 
one would expect that sterile neutrinos also form generations. 
If one such generation were discovered, it would strengthen confidence in the existence of others. 
The discovery of sterile neutrinos could give an answer to a variety of fundamental questions in particle physics (explanation of the nature of active neutrinos oscillations, the structure of neutrino mass matrix, an extension of the Standard Model, a non-conservation of lepton number), astrophysics and cosmology (dark matter). 
Upper limits for the sterile neutrino mixing matrix would give restrictions on the parameters of possible extensions of the Standard Model.

A great effort of the international scientific community is directed toward investigation of neutrino oscillations. For details see recent reviews~\cite{Feldman:2013vca,Nakaya:2015ksa}. However, the information obtained in oscillation experiments gives only the difference of mass squares of species involved and does not provide the absolute mass values. 
And neutrino oscillations can provide us with information about light sterile neutrinos only, with the mass of order eV. 
Only direct methods could allow to study the region of a higher masses.
In the keV mass range the information on possible existence of sterile neutrinos is obtained in precise measurements of $\beta$-decay spectra of various radioactive elements 
or in measurements of nuclear recoil in $K$-capture. 

The neutrino flavor states, such as $\nu_e$, $\nu_\mu$, $\nu_\tau$ and sterile neutrinos $\nu_s$ are not mass eigenstates, and can be represented as coherent sums of such states: $\mid\nu_\alpha\rangle=\sum\limits_i U_{\alpha i}\mid\nu_i\rangle$, 
where $\nu_\alpha$ and $\nu_i$ are flavor and mass states correspondingly and $U_{\alpha i}$ is the mixing matrix. 
Then the spectrum of electrons in beta decay can be represented~\cite{Klapdor} as $S(E)=\sum\limits_i U_{e i}^2  S(E,m_i^2)$ 
where $S(E,m_i^2)$ is the spectrum with a specific neutrino mass. 
Let $E_0$ corresponds to the end point of the spectrum at $m = 0$. 
For a non-zero $m$ the end point of $S(E,m_i^2)$ is given by $E_i = E_0 - m_i$, and at $E > E_i$ the eigenstate "i" does not contribute. 
The number of sterile neutrino states is unknown. 
Let $m_N $ correspond to the lightest neutrino among the "heavy" states, $m_N \gg 1$ eV. When searching for such neutrino the masses of active neutrinos (or a possible light sterile neutrino) can be set to zero. 
Therefore one can get a simple expression for the shape of the electron spectrum in beta decay
\begin{equation}
{S(E) =  U^2_{e}S(E,m_N^2)+(1-U^2_{e})S(E,0),}
\label{eq:funct}
\end{equation}
in the energy range $E > E_0 - m_X$ even if a heavier neutrino exists, $m_X \gg m_N$. 

\begin{figure}[tbh]
\center
	\includegraphics[width=0.8\linewidth]{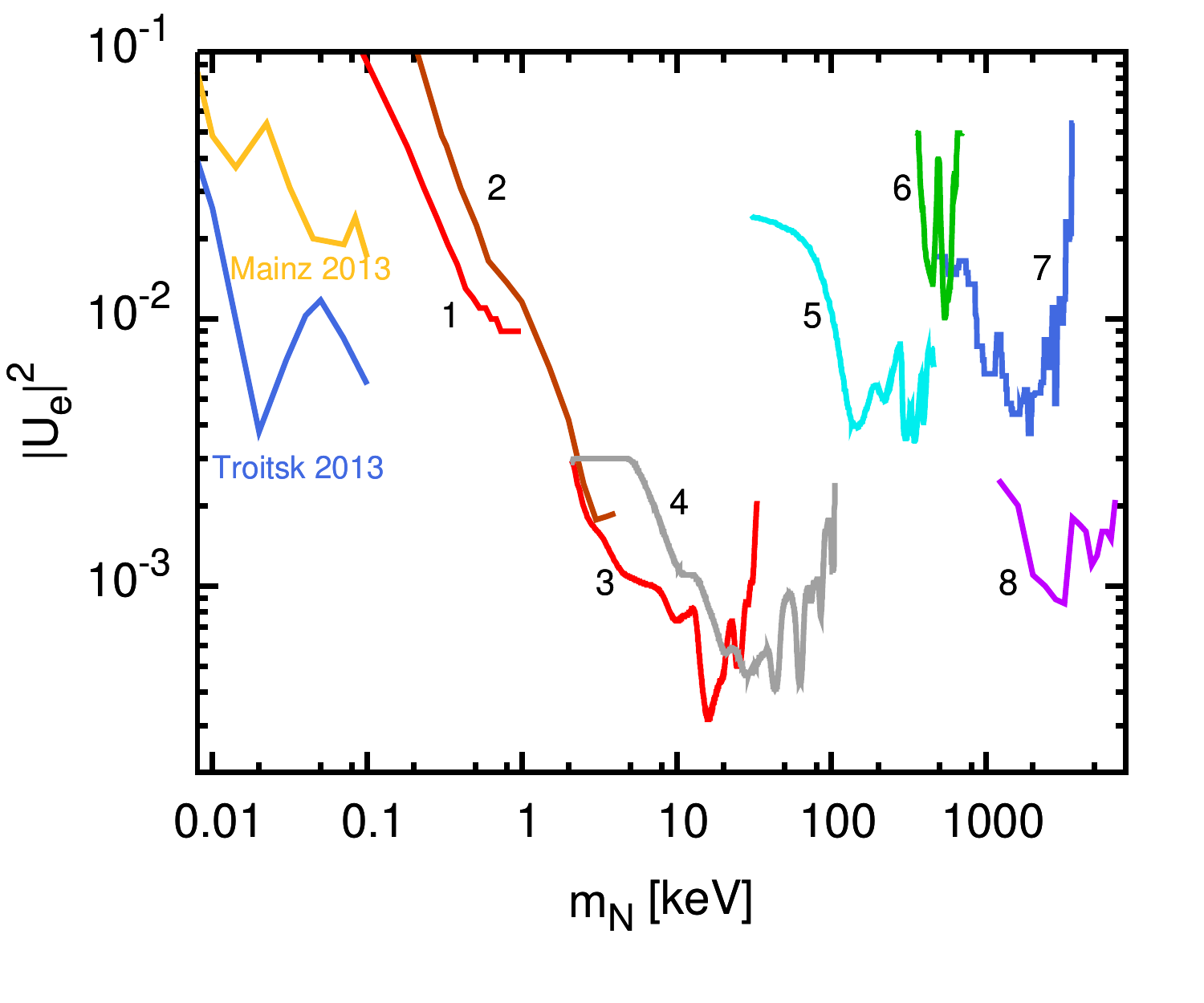} 
	\caption{Current limits on sterile neutrinos in radioactive decays. Curves numbered from 1 to 8 correspond to Refs.~\cite{Galeazzi:2001py,Hiddemann:1995ce,Holzschuh:1999vy,Holzschuh:2000nj,Schreckenbach:1983cg,Hindi:1998ym,Trinczek:2003zz,Deutsch:1990ut}, ''Mainz 2012'' is from~\cite{Kraus:2012he}, and ''Troitsk 2013'' - from~\cite{Belesev:2012hx,Belesev:2013cba}. }
	\label{limits_now}
\end{figure}

The current best limits on $U^2_{e}$, obtained in the past direct experiments,  are shown in Fig.~\ref{limits_now}. 
In the mass range $m_N < 0.1$ keV the best limits were reported recently by Mainz~\cite{Kraus:2012he} and Troitsk~\cite{Belesev:2012hx,Belesev:2013cba}.\footnote{The best direct limits on the electron antineutrino mass from tritium $\beta$-decay endpoint kinematics have been obtained in the past by these two experiments ~\cite{Aseev:2011dq,Kraus:2004zw} as well. 
This program, including extension for the search of keV-scale sterile neutrino, will be continued at KATRIN experiment~\cite{Angrik:2005ep,Mertens:2014nha}.} 
In Troitsk, we are planning to extend these searches and to probe sterile neutrino mass range up to a few keV. 
This paper studies sensitivity of the existing Troitsk experiment in the corresponding mass range and considers upgrades needed to improve the sensitivity.

The rest of the paper is organized as follows.  First we present technical details for major components of the installation. Then we discuss the current systematic errors that limit our sensitivity. 
Finally, we briefly present plans for an upgrade for the nearest future.

\section{The  Troitsk nu-mass experiment}

The "Troitsk nu-mass" program is conducted by the Institute for Nuclear Research of the Russian Academy of Sciences. It was initiated by Vladimir Lobashev with a goal to limit the mass of electron anti-neutrino by analyzing the shape of tritium beta decay  spectrum  near the end point~\cite{Lobashev:2003kt}. The corresponding measurements were done from 1985 to 2005~\cite{Belesev:1995sb,Lobashev:1999tp} and the final results of these efforts were published in~\cite{Aseev:2011dq}. Later on these data were re-analyised in a search of  admixture of sterile neutrinos with masses from 3 to 100 eV~\cite{Belesev:2012hx,Belesev:2013cba}. The interval of masses has been limited by the energy range measured in the past, which was about 200 eV from the spectrum endpoint (excluding monitor points). Currently we are expanding the energy range of measurements up to 5 keV from the endpoint with a goal to probe sterile neutrinos with masses up to 4 keV. 

At the first stage of the new program the "Troitsk nu-mass" experiment will utilize the same layout as in the past measurements~\cite{Aseev:2011dq} but with a new spectrometer. The two main components of "Troitsk nu-mass" experiment are the Windowless Gaseous Tritium Source (WGTS) and the Electrostatic Spectrometer with Magnetic Adiabatic Collimation (ESMAC). Gaseous source, while being very complex to manufacture and maintain, is needed to avoid solid state effects such as source charging, neighbor excitation and backscattering from substrate~\cite{Kraus:2004zw}. "Windowless" means that the radioactive gas freely circulates in a source and there are no effects associated with a wall.
Working principle of ESMAC is based on the electrostatic rejection of all electrons with energy less than a given spectrometer potential. All electrons with higher energy are counted by the detector thus giving an integral electron energy spectrum. In this procedure all electrons have to be aligned (adiabatically collimated by magnetic field) at the spectrometer analyzing plane. The operation of the spectrometer requires the adiabatic transport system with magnetic fields up to 8 Tesla which in turn requires  superconducting magnets and cryogenic systems with circulating liquid helium.

Recently, the main spectrometer has been replaced with a larger vessel. The new one is about two times longer, it has the twice-larger diameter and, as the result, almost the ten times larger volume. The new main magnet has also a stronger by about 20\% magnetic field, Fig.~\ref{fig:Spectrometer_Bz}. The longitudinal magnetic field in the center of the spectrometer has a magnitude lower by a factor two compared to the old spectrometer. As the result of all these factors, following the Eq.~\ref{eq:resol},  the energy resolution has been improved to 1.5 eV, which is more than a factor of two better than for the old spectrometer.

\begin{figure}[ht]
   \includegraphics[width=0.95\linewidth]{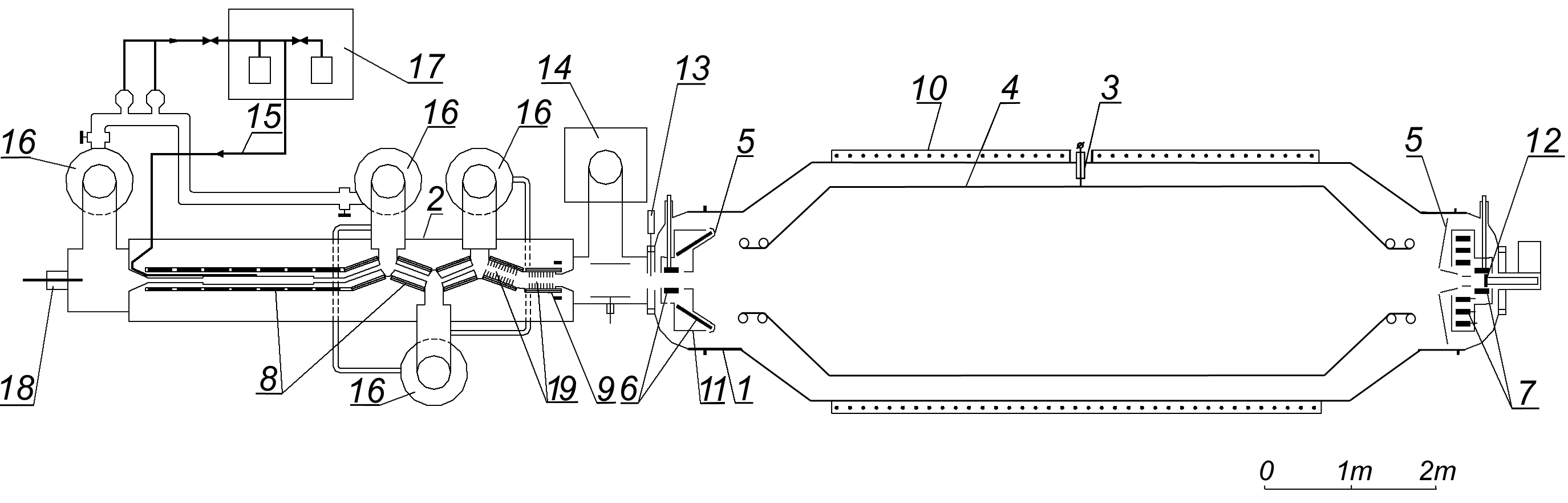}
   \caption{The current layout of "Troitsk nu-mass" experiment. WGTS and its pumping system is on the left side, spectrometer -- on the right. 1 -- Spectrometer vacuum tank; 2 --  Source vacuum tank; 3 --  High voltage feedthrough; 4 --  Spectrometer electrode;  5 --  Ground electrodes; 6, 7, 8, 9 --  Superconducting coils; 10 --  warm coils; 11 -- L-N$_2$ jacket; 12 --  L-N$_2$ cooled Si(Li) detector; 13 --  Fast shutter; 14 --  Ti-pump ; 15 --  T$_2$ Loop; 16 --  Diffusion pumps; 17 --  T$_2$ storage and purification system; 18 --  Electron gun; 19 --  cryo Argon pump.}
   \label{fig:numass_layout}
\end{figure}

\subsection{Windowless Gaseous Tritium Source}

Electrons from the $\beta-$decay of molecular tritium are produced in WGTS, which is a pipe of 3 m long and  5 cm in diameter. Both sides of the pipe are ended by transport channels with a diameter of 2 cm and a length of 1 m each. 
During measurements the tritium is injected into the center of the tritium pipe through a thin capillary. 
Maximum density of the working gas reaches $ 5 \cdot 10^{14}  \rm ~cm^{-3} $. 
On the rear side of WGTS a mercury diffusion pump is located. 
Tritium gas is pumped towards spectrometer by a set of another three sequential mercury diffusion pumps along zig-zag transport channels. This shape prevents molecules of the gas to enter  the spectrometer. All four pumps direct gas into collector, then another small booster mercury diffusion pump return gas  into WGTS. In this way a closed loop is formed for tritium circulation.

The tritium pipe and transport channels are located inside cryostats with a superconducting magnets for providing transport field. In order to prevent gaseous tritium freezing on the walls, the pipe is isolated from the cryostats and heated continuously with helium gas at 30 K. This setup provides also thermal stabilization of tritium in the source. This is needed because temperature variations by $\pm$1 K at  30 K  will change the gas density in WGTS by $ 3\% $. As a result of thermal stabilization, the density fluctuations are maintained at 1\% level. This value is required to be less than the precision of 3\% with which we measure the column density in WGTS with electron gun. 

\begin{figure}[ht]
	\includegraphics[width=0.95\linewidth]{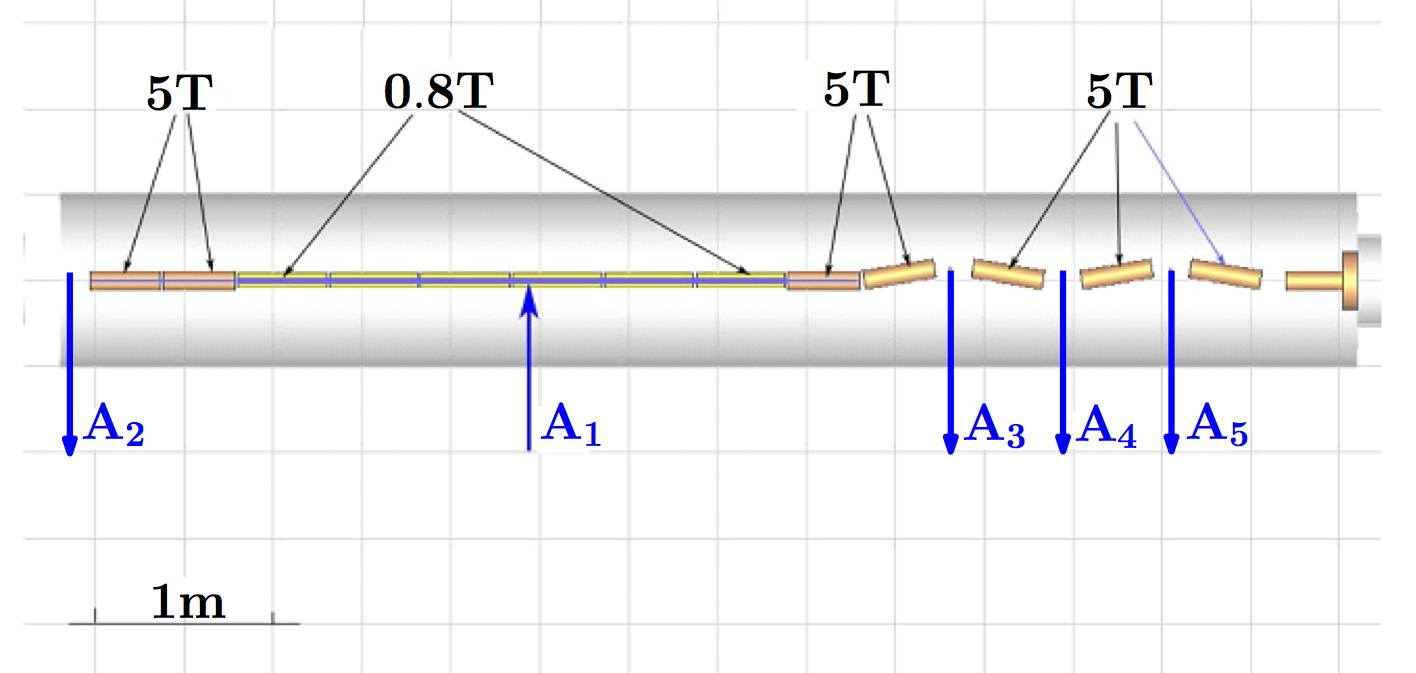} 
	\caption{Location of superconducting magnets with axial field in WGTS and in transport channel. The strength of magnetic field in Tesla is marked on the top. Arrow labelled by $A_1$ shows the injection point of the gas, $A_2 - A_5$ show positions of mercury pumps.}
	\label{fig:magnets}
\end{figure}

A set of superconducting coils is positioned around WGTS and transport system, Fig.~\ref{fig:magnets}. Electrons are winding around axial magnetic lines and move to the rear and front sides (i.e. downstream and upstream sides) of WGTS.  Upstream electrons within spectrometer acceptance angle are transported to the spectrometer, while the rest is rejected by the pinch magnet.  Downstream electrons  move to the rear wall. Electrons with very large angle with respect to the magnetic field direction are trapped inside WGTS. Majority of those electrons settle on the walls eventually by colliding with the gas. A small portion (10$^{-4}$) of trapped electrons may get into the spectrometer and distorts the integral spectrum. In details trapping effect is discussed in the section 3. 

The last evacuating element in front of the spectrometer is the cryo-pump with frozen argon which effectively absorbs possible remaining tritium.

\subsection{Spectrometer}

\begin{figure}[htb]
	\includegraphics[width=1.\linewidth]{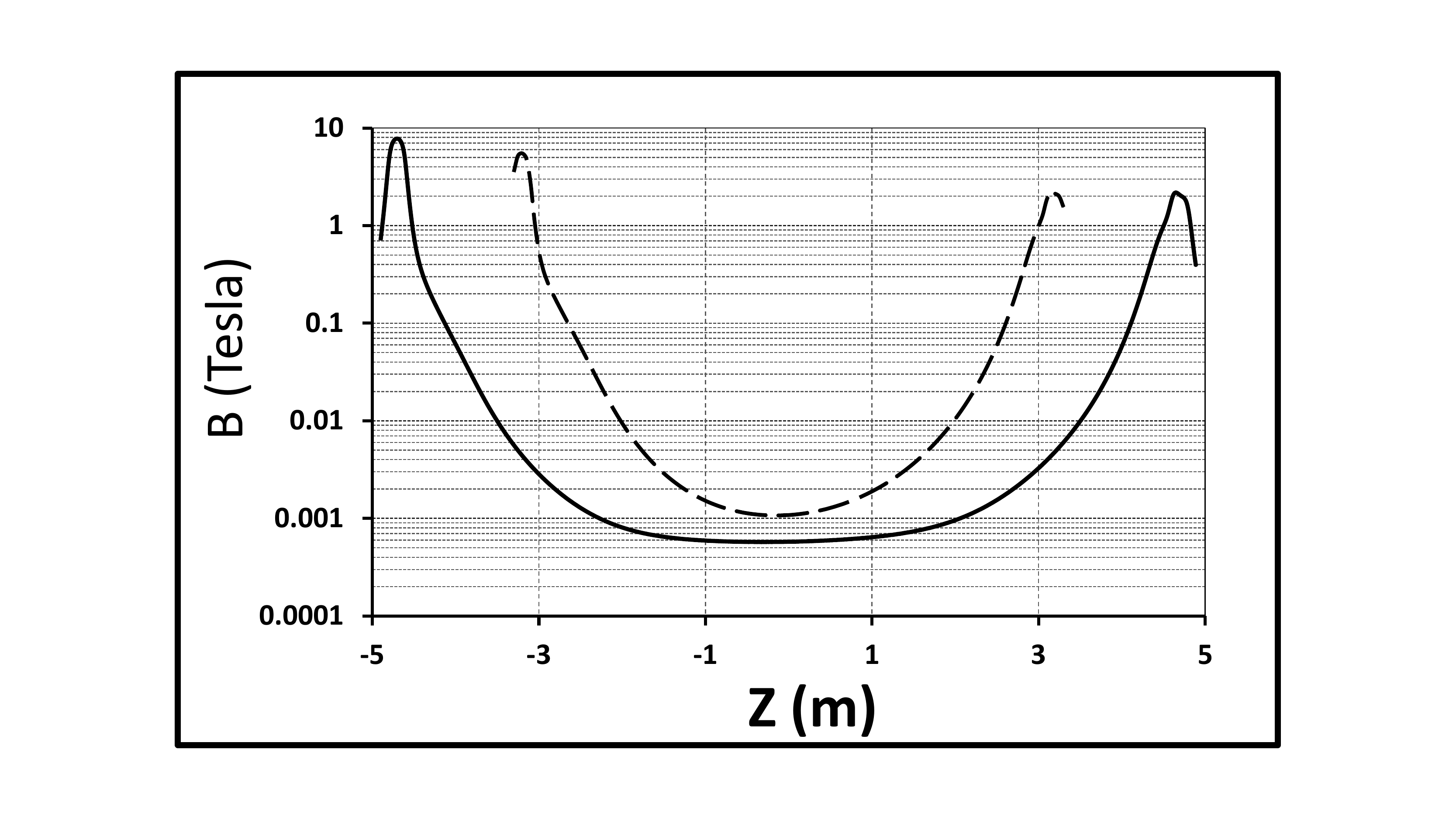} 
	\caption{Magnetic field profile at the spectrometer z-axis. For comparison, dashed line presents magnetil field for the old spectrometer. Pinch magnet is located on the left, detector is positioned inside 3T magnet on the right. }
	\label{fig:Spectrometer_Bz}
\end{figure}

The energy spectrum of $\beta$-electrons is measured in the spectrometer by varying the electrostatic potential which is applied to it. Electrons with energy smaller then this potential are reflected. Electrons with higher energy pass the electrostatic  barrier and hit the detector. In this way, the integral spectrum of electrons is measured. However, the retarding potential changes only the longitudinal component of the energy, $ E_{||} \propto v_{||}^2$. Therefore, in order to get good energy resolution one has to make the transversal energy of electrons, $ E_{\perp} \propto v_{\perp}^2$, to be as small as possible at the potential barrier, i.e. in the analyzing plane of the spectrometer.
This is achieved by a special configuration of induced magnetic fields.

 The magnetic field lines of the spectrometer form a ``bottle'' shape with  highest longitudinal field at the pinch-magnet located near spectrometer entrance~\cite{Lobashev:1985mu}. The field in the pinch-magnet, $B_0$, is up to 8 Tesla. The lowest longitudinal magnetic field is in the analyzing plane in the center of spectrometer (Fig.~\ref{fig:Spectrometer_Bz}). The magnetic moment of a charged particle, $\mu=E_{\perp}/2B$, is an adiabatic invariant when the transversal gradient of the magnetic field is small. We have in this case $ E_{\perp m} = E_{\perp 0} \cdot B_m/B_0$,  where subscripts $m$ and $0$ refer to quantities in analyzing mid-plane and in pinch-magnet correspondingly. Therefore, resolution of the spectrometer (spread of  transversal energies of electrons) is: 
\begin{equation}
{\Delta E _{\perp m} = E_{0} \cdot B_m/B_0,}
\label{eq:resol}
\end{equation}
which boils down to $\approx$ 1.5 eV at highest energies, $E_0 \approx 18$ keV. This simple expression neglects spatial distribution of the electrostatic potential, however. More details and accurate formula are presented in \cite{Aseev:2011dq}.

\begin{figure}[htb]
	\includegraphics[width=0.9\linewidth]{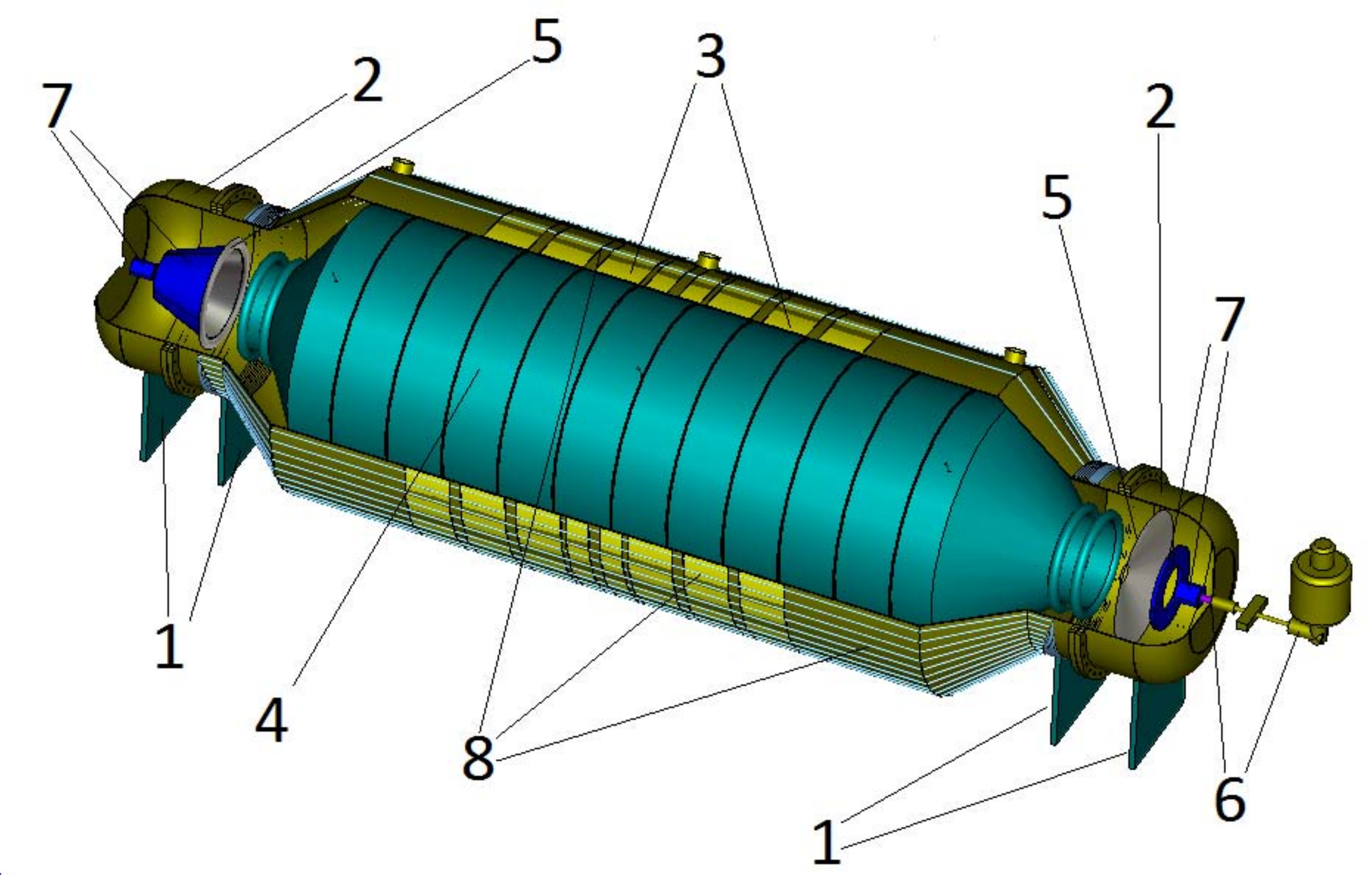} 
	\caption{General view of the spectrometer. 1-  supports, 2 – side cups, 3 – axial winding, 4 – main high voltage electrode, 5 – additional ground electrodes, 6 – detector with liquid $N_2$ dewar, 7 - superconducting solenoids, 8 –  correction coils}
	\label{fig:spectrometer}
\end{figure}

The spectrometer itself consists of a central vessel and two removable side caps,  Fig.~\ref{fig:spectrometer}.  The vessel is a stainless steel cylinder with a conical taper at the edges for providing a transition from the diameter of the  cup to the diameter of the housing central part. Sectional high voltage  electrode is kept under common high voltage and is mounted on insulators. To provide mechanical rigidity, the strengthening ribs are welded  on the outside of the central part of the spectrometer. Cups are hemispherical metal structures with attached pumping ports, inputs and outputs of the cryogenic system, the current leads for superconducting magnets and vacuum inputs and outputs for the sensor cables and also contain  superconducting magnets inside.

Magnetic field in the spectrometer is formed by a system of two main electromagnets. Superconducting cryogenic magnets produce a field of up to 8 T at the input of the spectrometer (pinch-magnet) and up to 3 T at the detection side. Magnets are rather small, so the field quickly decreases with distance from the coils  and in the center of the spectrometer reaches  3 Gs. Fine tuning of the field to the desired magnetic shape  in the central part of the spectrometer is controlled by warm electromagnetic coils winded outside the spectrometer vessel.  These four coils can generate an additional axial field up to 4 Gs in the central analyzing plane of the spectrometer. In Fig.~\ref{fig:Spectrometer_Bz} the longitudinal component of magnetic field is shown at as a function of distance from spectrometer analyzing plane. The Earth's magnetic field and other external fields are compensated by two  warm transverse coils which can form 1.2 Gs field  in the transverse vertical and horizontal directions. Currents in these coils on the level of few Amperes were adjusted to get the maximum  rate for electrons from WGTS.

Spectrometer is equipped with a heating system which is used for out-gassing of the walls during pre-run maintenance. During the run the spectrometer is pumped off by two magnetic-discharge pumps. Superconducting coils at the side ends of the spectrometer also work as a cryo-pumps. The pressure of the gas in the spectrometer does not exceed $ 5 \cdot 10^{-9}  \rm ~mbar $. High vacuum is needed, firstly, to block radioactive gas away from the spectrometer (decays in spectrometer create a non-uniform background), and, secondly, to ensure the passage of electrons through spectrometer without scattering.

\subsection{Detector}

\begin{figure}[htb]
\includegraphics[scale=0.5]{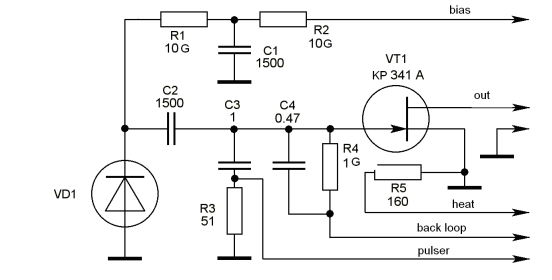}
\caption{\label{fig:amplif} The head cascade of the preamplifier which is cooled by liquid nitrogen and heated by an additional resistor R5. Capacitor units are pF for C1-C3 and $\mu$F for C4.}
\end{figure}

\begin{figure}[htb]
\begin{center}
	\includegraphics[width=0.5\linewidth]{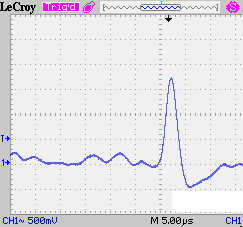}
        \caption{A typical signal at the output of the shaper for 18 keV electron.}
	\label{fig:signal}
\end{center}
\end{figure}

The current detector is based on custom made Si(Li) p-i-n  diode, 25 mm in diameter, with thin gold plated 20 $\mu \rm g/cm^2$ entrance window. The entrance window is a source of a 4 keV threshold. The detector aperture is limited by a copper collimator, 17 mm in diameter. The signal is amplified by a charge sensitive preamp. The first cascade of preamp, Fig.~\ref{fig:amplif},  is  located in a vacuum pipe one meter apart from the main cascades. Detector and this vacuum pipe are positioned inside the last superconducting solenoid  of the spectrometer and are cooled down by the liquid nitrogen. The optimum operation temperature for the field transistor used is about 120 K, therefore it is heated up by an additional resistor, see Fig.~\ref{fig:amplif}. The noise amplitude goes down with temperature $T$ as $\sqrt T$ but at temperature below 100 K transistor stops working. To get the best signal to noise ratio, current through this resistor is controlled by an additional voltage power supply. The current  was optimized at the very beginning and was kept at this value for the rest of the measurements. Such a cooling gives about factor 1.5 better signal to noise ratio. The signal from  preamplifier is then amplified and shaped with shaping time of 1 $\mu$sec. The detector readout is triggered when the  signal exceeds some threshold equivalent to 4 keV electrons. An example of  signal from 18 keV electron which is digitized  is shown in Fig.~\ref{fig:signal}. The average conversion factor between electron energy and signal amplitude is about $0.1~V/keV$.

The signal is digitized by 4096 channel amplitude to digital converter. The maximum counting rate in previous measurements did not exceed 5-10 kHz with constant dead time of 7.2~$\mu$sec. To search for sterile neutrino signal the counting  will be increased to 20-30 kHz with the current electronics. The nearest future plan is to do 100-120 MHz signal sampling.

We are currently working on two options for the detector system upgrade. 
The first option is to use a bigger size segmented Si(Li) detector with 12-16 channels. 
The second option is to use an array of windowless multi-pixel avalanche photo diodes (M-APD). 

\subsection{Photoelectron gun}

\begin{figure}[htb]
	\includegraphics[width=0.95\linewidth]{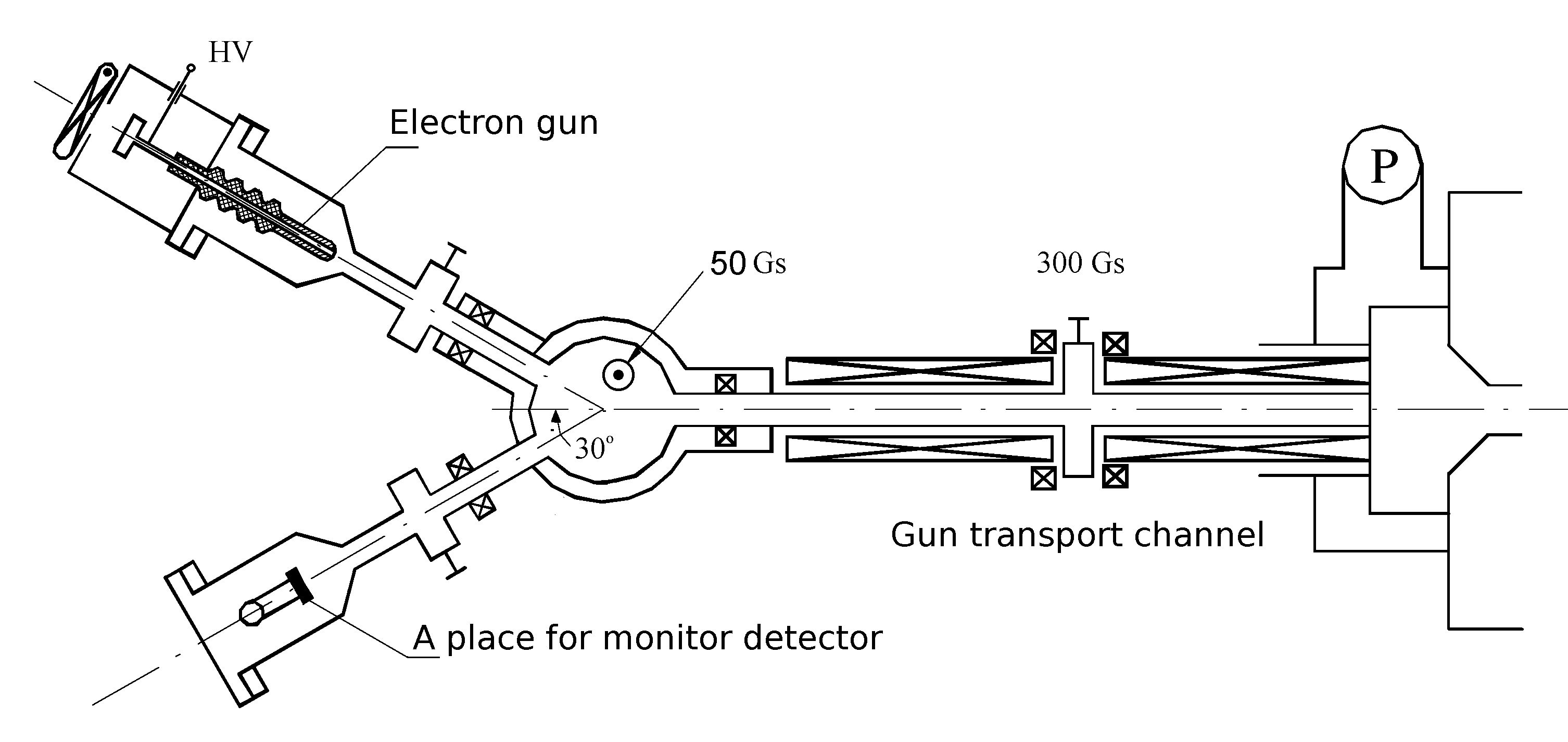}
        \caption{Location of the electron gun and place for monitor detector (currently not used) at the rear side of WGTS.}
	\label{fig:gun}
\end{figure}

We make necessary calibrations of spectrometer and detector systems using photo-electron gun located at the rear side of the WGTS. 
The ultraviolet deuterium lamp is illuminating  through the quartz window onto the quartz half-transparent gold cathode, which is under negative potential. As a result, photoelectrons escape gold surface with energy of the applied electrostatic potential. Electron energy spectrum and its spread are defined by calibration and stability of the high voltage system. Energy resolution is better than 0.3 eV. The current intensity stability depends on stability of deuterium lamp and is less than 0.5\% for one calibration session of about 0.5 hour. Photoelectrons produced in the gun are delivered to the first superconducting magnet of the WGTS by the electron gun transport system consisting of two dipole magnetic lenses, bending magnet and adiabatic transport channel, see Fig. 8. Bending magnet serves for three purposes:

-- To avoid direct illumination of WGTS by the gun's ultraviolet light.

-- To avoid bombardment of the gun window by accelerated positive ions produced by ionization in WGTS. 

-- To reserve place for monitor detector, which is not currently used.   Electrons, which are produced  in WGTS by tritium decay to the rear side direction, will move back through the transport system, deflect to the left by the bending magnet and hit the monitor detector.

The adiabatic transport channel provides a smooth change of the magnetic field from 300 gauss to 5 Tesla at WGTS entrance.

\subsection{HV}

\begin{figure}[htb]
\includegraphics[scale=0.4]{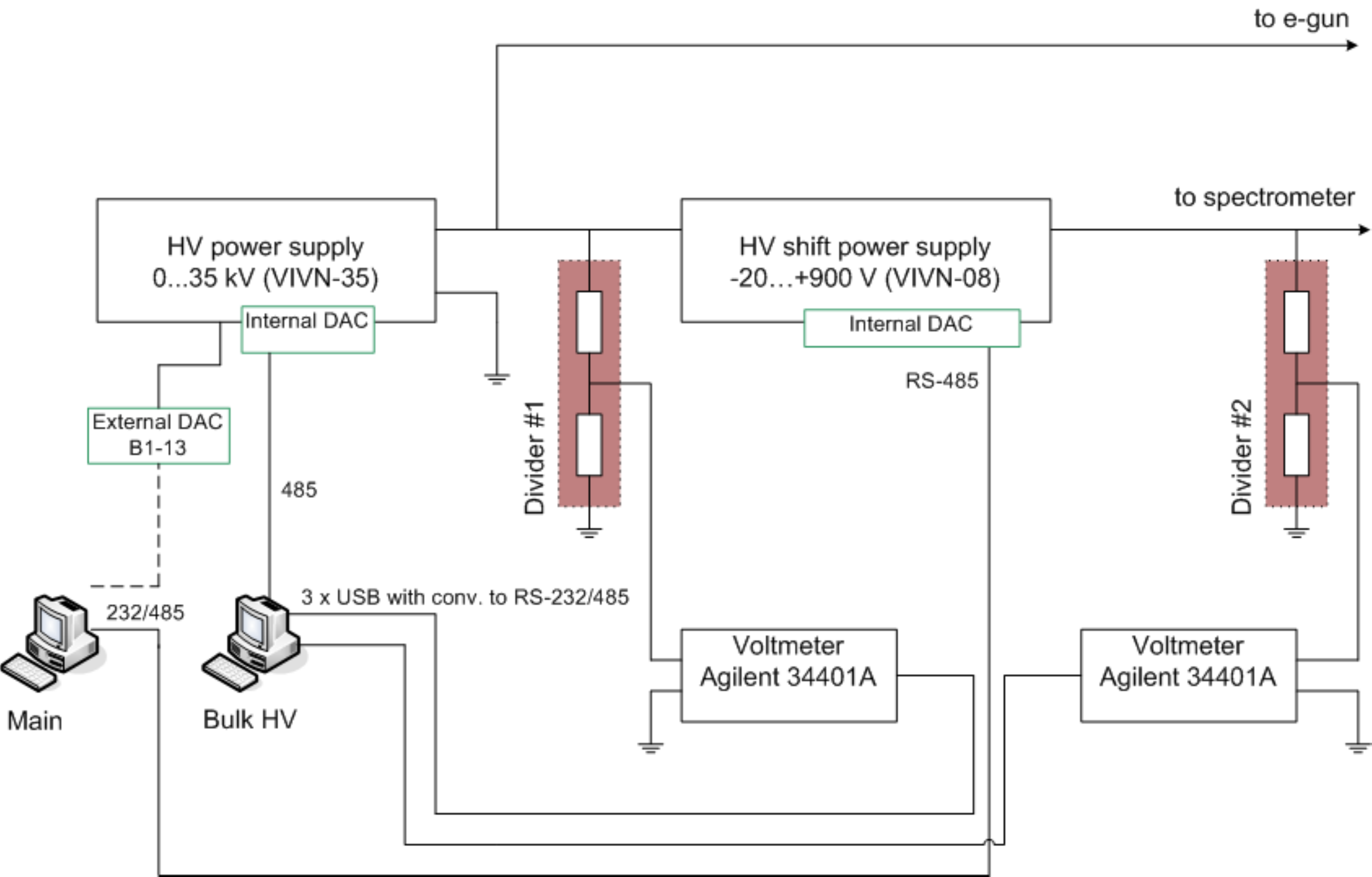}
\caption{\label{fig:hv} High Voltage system. The bulk power supply up to 35 kV (VIVN-35) is connected to the entrance of the other power supply which works as a HV shifter.}
\end{figure}

Schematic diagram of the current High Voltage (HV) system is presented in Fig.~\ref{fig:hv}. The main power source up to -35 kV (VIVN-35, Russia) is chained with another HV supply (VIVN-08), the latter serves as a high precision shifter in a small potential range from +900V to -20V on top of the main HV. During calibrations, this system  fixes  energy of electrons from the gun at the level of main HV, and allows to vary additionally the potential on the spectrometer within the range of the shifter. This configuration allows to avoid random relative drift of both potentials. Voltages are controlled by two high precision voltage dividers and voltmeters Agilent 34401A with 6 $1/2$ digits. During actual measurements the HV shift is set to zero.

\subsection{Cryogenics}

Superconducting magnets extend to a distance of  20 meters and are cooled down by  circulating two-phase helium at  overpressure of 0.5-0.8 bar. Two-phase mixture at temperature 4.5-4.6 K contains $~$40\% liquid and $~$60\% vapor helium. The mixture passes sequentially through 12 cryostats and returns to 10 liters cryostat where the gas fraction at temperature 4.6 K is separated and sent back to the {\bf  Linde} TCF50  helium refrigeration system. Effective volume of all cryostats was minimized and the system uses 100 liters of liquid helium  in total.

\subsection{Software}

Data taking system could be split  on three parts. The first one accumulates data via a set of CAMAC modules, manipulates high voltage settings. On-line monitoring allows to display and control counting rates, status and values of HV. The other part of software serves as a slow control system. It writes values of all vacuum sensors and temperatures in WGTS and  spectrometer. It also continuously monitor the composition of  the gas in WGTS through the mass analyzer. The third part of software allows to operate and control the helium refrigeration system and other elements of  cryogenics and is rather independent. It sets alarms and stops the refrigeration process in case of some problem.

\section{Experimental sensitivity}

The sensitivity to the mixing angle of sterile neutrino in the keV mass range at the first stage of our experiment (i.e. on the existing equipment) is limited by systematic errors only. 
This is because the data required to reach a comparable statistical sensitivity can be acquired in a month of pure measurement time, which corresponds to four experimental runs. 
In this section we describe sources of the known systematics. 
The systematical errors are calculated by a numerical modeling of the spectrum distortions caused by changes of the corresponding systematical parameters within their boundaries.

The major systematic limitations at "Troitsk nu-mass" are related to:
\begin{itemize}
\item an insufficiently accurate determination of the electron energy losses in the source of molecular tritium and the related fluctuations of the gas column density;
\item the electronics dead time and pile-up uncertainty;
\item an insufficient high voltage stability;
\item spectrum distortions caused by the deviation of the transmission function of the spectrometer from the estimated one or by an inaccurate account of the energy dependence of the registration efficiency;
\item the trapping of electrons in WGTS.
\end{itemize}
In what follows we discuss these errors in details. The procedure we used to estimate expected sensitivity is similar to the one implemented in \cite{Aseev:2011dq}. In final analysis more advanced techniques from \cite{Belesev:2013cba} will be used.

\begin{figure}[htb]
	\includegraphics[width=0.8\textwidth]{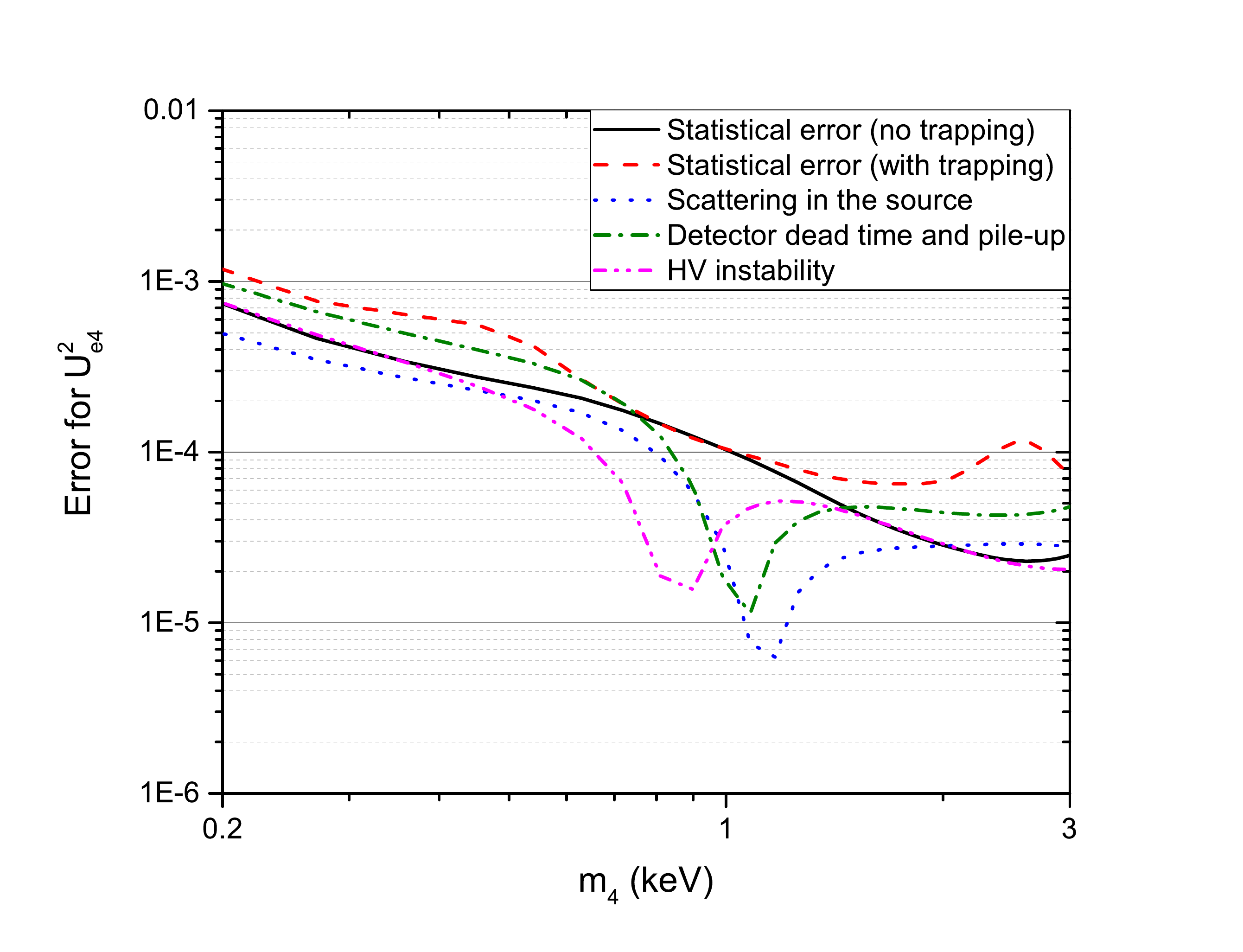}
	\caption{The impact of systematical and statistical errors on the sensitivity to the mixing parameter of sterile neutrinos. The statistical error corresponds to one month of data acquisition.}
	\label{fig:syst_err}
\end{figure}

\paragraph{Thickness of the source and final states spectrum.}

A major uncertainty of the electron energy losses is related to the uncertainty in the gas column density. 
The latter can be characterized by the effective thickness, $X$, which is the ratio of the distance traveled by electrons to the mean free path. 
It can be measured "on-line" in our experiment by the electron gun with the accuracy better than 3 \%. 
Therefore, when  $X =0.3$, we know it with an absolute error approximately equal to $0.01$, which in turn translates into the error on $U_e$ shown in Fig.~\ref{fig:syst_err} by the dashed line. 
This value, $X =0.3$, was typical for our runs in the past in the spectrum measurements near the endpoint that yielded the limits on electron neutrino mass. 
Now, for measurements away from the endpoint, the source intensity can be lowered. 
This will result in a lower $X$ and a lower error on $U_e$. 
Therefore, the estimate shown in Fig.~\ref{fig:syst_err} should be considered as conservative.

\paragraph{Transmission function and detection efficiency.}

New challenges, which we will face in the new experiment, are related to the operation of the spectrometer in a wide energy range. 
When the energy of an electron entering the spectrometer is significantly higher than the retarding potential, its velocity in analyzing plane of the spectrometer can not be considered small. 
For a large difference between the electron energy and the retarding potential of the spectrometer,  the transmission function of the spectrometer may not be constant. Simulations show that distortions occur only at the difference $ E - U > $ 5 keV. Therefore, it will not affect our measurements during the first phase when $\beta$-spectrum will be measured in the 14--19 keV energy range. 
Subsequently, this effect, as well as the detection efficiency (change of detector sensitivity for different electron energies), will be studied experimentally with the electron gun. 
Our estimates show that the ultraviolet light source D-2000-DUV (Ocean Optics) with the declared short time stability of $<0.005\%$ and an optical drift of less than $0.5\%$ per hour has a sufficient intensity stability for our calibrations.

\paragraph{Dead time and pileup.}

The dead time of the detector readout and the event pile-up are crucial even at  moderate electron count rates $\sim 10^4$ Hz. The current data acquisition system has fixed dead time 7.2 $\mu$s with uncertainty 0.03 $\mu$s. For the first phase of planned experiment we will make an upgrade of detector and data acquisition system  to obtain dead time 0.1 $\mu$s with  uncertainty better than 0.01 $\mu$s.  Corresponding error on $U_e$ is shown in Fig.~\ref{fig:syst_err} by the dotted line.

\paragraph{Trapping.}

Another important effect associated with our tritium source is, so-called, trapping effect. Only a  fraction of electrons from WGTS enters spectrometer, the rest of electrons is locked in a magnetic trap. Majority of those electrons lose energy in collisions with gas and settle on the  walls eventually. However, a small portion (10$^{-4}$) of trapped electrons may get into spectrometer at some point of the evolution. The energy distribution of those electrons is significantly different from the distribution of electrons entering the spectrometer directly. Although the spectrum shape of the trapping effect is well reconstructed in numerical modeling, its amplitude does not. Our simulations have shown that the amplitude of trapping effect can be included as a free parameter in the analysis of the spectrum using 5-parameter model (normalization, background, endpoint energy, heavy neutrino contribution and trapping) instead of 4 parameters. We also found that the correlation between trapping amplitude and heavy neutrino contribution is relatively small. These parameters can be disentangled, meaning that the value for trapping amplitude and $U^2_e$  could be obtained from the very same experimental data without any additional assumptions.

The resulting expected sensitivity at 95\% C.L to the mixing parameter $U^2_e$, which can be achieved at the existing equipment with minimal modifications, is shown in Fig.~\ref{expect} by the dash-dotted curve. 
All known systematics discussed above and displayed in Fig.~\ref{fig:syst_err} have been taken into account while the statistics corresponds to 30 days of data acquisition with the integral source intensity of $\approx 5\cdot10^6$ decays per second (which corresponds to the maximum counting rate $\approx 10^5$ at 14 kV).

\section{Upcoming upgrades}

\begin{figure}[tbh]
\center
	\includegraphics[width=0.8\linewidth]{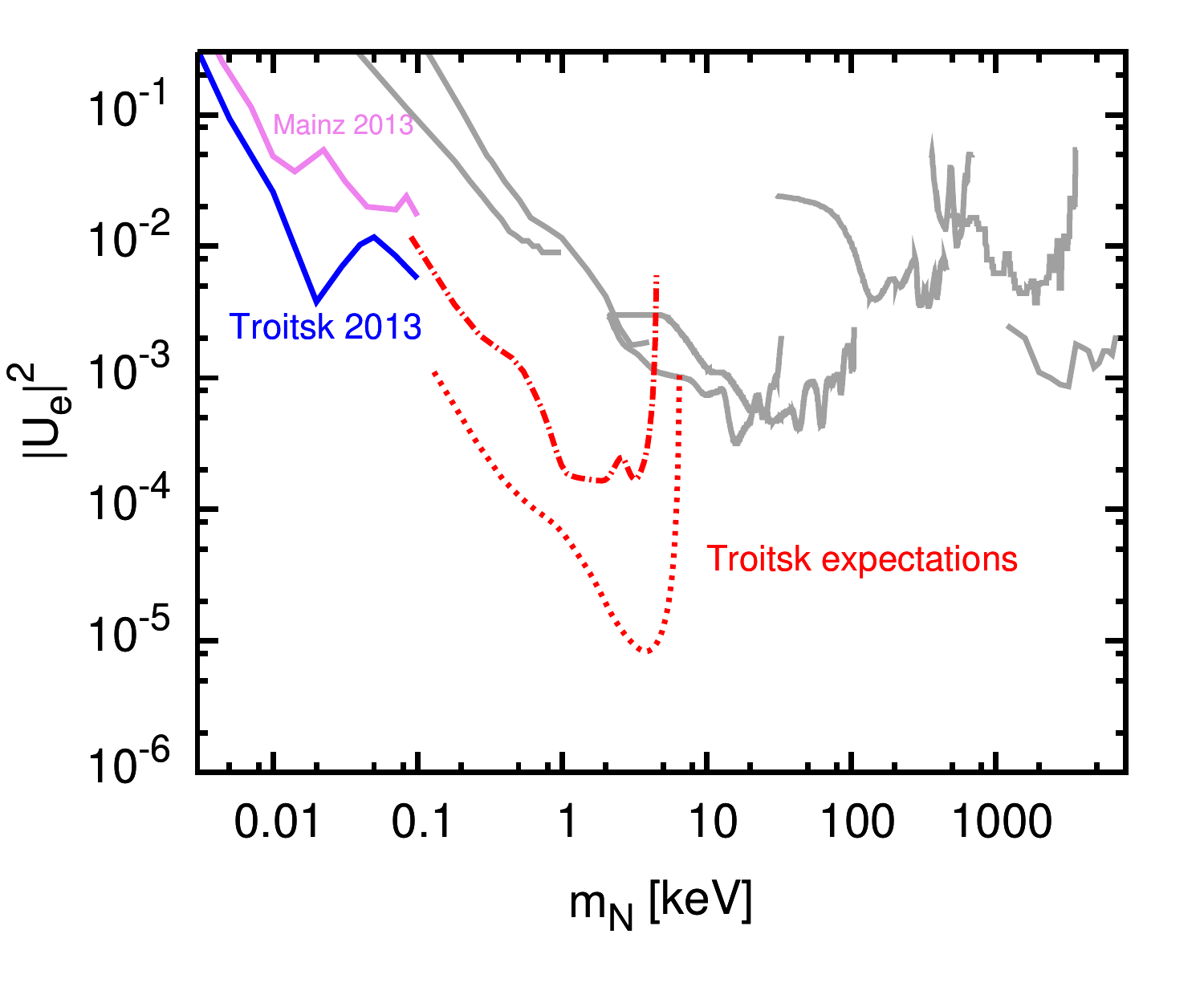} 
	\caption{An estimate of sensitivity to the presence of additional sterile neutrinos depending on the neutrino mass at "Troitsk nu-mass". 
	The upper dash-dotted curve corresponds to measurements at the existing equipment, 
	the lower dotted curve, after the upgrades. 
	The gray lines are the existing limits from Fig.~\ref{limits_now}.}
	\label{expect}
\end{figure}

When the measurement at the level described in previous section will be completed, 
upgrades necessary to reach a better sensitivity are planned. In particular, there is a need for better calibrations with electron gun, which could be achieved with a new deuterium lamp, and more precise knowledge of readout dead time and pule-up with a new fast digitizing electronics. The required calibrations are as following: 
\begin{itemize}
\item the relative long-term stability of intensity of the electron gun should be better than 0.1\% over one hour; 
\item the precise measurements of spectrometer transmission function in a wide energy range with electron gun;
\item the absolute accuracy of determination of the effective thickness of the source will be of the order or better than 0.001 in units of the inelastic scattering probability. This calibration should be done before and after each run, that is, with a period of less than one hour; 
\item the accuracy of determination of the effective dead time of the data acquisition system will be of the order of a few ns (at rates up to 200 kHz). 
\end{itemize}
In addotion, some other experimental conditions will be optimized:
\begin{itemize}
\item the intensity of the tritium source should be increased to get a better statistics without loss of sensitivity at high rates.
\item the absolute stability of the high voltage system will be better than 0.1~V;
\end{itemize}
The expected limits are approximately two orders of magnitude stronger than the existing ones, see the lower dotted curve in Fig.~\ref{expect}.

Finally, in order to extend the range of probed neutrino masses, a new experiment is currently in preparation in Troitsk~\cite{Abdurashitov:2014vqa}. 
This new experiment will be employing conventional technique of proportional counters filled with portion of tritium gas, 
but with an up-to-date fast signal digitization at high count rates, up to $10^6$ Hz.

\bibliographystyle{JHEP}

\bibliography{LOI}

\end{document}